\documentclass[aps,notitlepage,twocolumn]{revtex4-1}
\usepackage{graphicx}
\usepackage{float}
\usepackage{amsmath}
\usepackage{amssymb}
\usepackage{hyperref}
\usepackage{bm}
\usepackage{color}
\usepackage{siunitx}
\usepackage{lineno}
\usepackage{orcidlink}
\newcommand{\thermal}[1]{\langle #1 \rangle}
\newcommand{\Tr}{\operatorname{Tr}}

\begin{document}

\title{Efficient simulation of a pair of dissipative qubits antiferromagnetically
coupled}

\author{Francesco G. Capone$^{1,2}$\orcidlink{0009-0000-9832-4385}}
\author{Giulio De Filippis$^{1,2}$\orcidlink{0000-0003-0557-3556}}
\author{Vittorio Cataudella$^{1,2}$\orcidlink{0000-0002-1835-1429}}
\author{Antonio de Candia$^{1,2}$\orcidlink{0000-0002-9869-1297}}

\affiliation{$^1$Dipartimento di Fisica ``E. Pancini'', Universit\`a di Napoli Federico
II,\\
Complesso Universitario di Monte Sant'Angelo, via Cintia, 80126 Napoli, Italy}
\affiliation{$^2$INFN, Sezione di Napoli, Napoli, Italy}

\begin{abstract}
We investigate the efficiency of different quantum Monte Carlo simulations of a pair
of antiferromagnetically coupled qubits in an Ohmic dissipative environment. Using
a Trotter-Suzuky decomposition and integrating out the degrees of freedom of the thermal
bath,
the model maps onto a frustrated long-range double-chain Ising lattice. We prove that:
i) due to frustration, the conventional Swendsen-Wang approach to cluster dynamics
turns out to suffer from a severe inefficiency, stemming from the
mismatch between spin correlations and cluster connectivity; 
ii) the Kandel-Domany approach is extremely effective in the study of dissipative
quantum qubits. We partition the double-chain into different types of plaquettes and
minimize the weight of graphs containing antiferromagnetic bonds by using both analytic
and numerical approaches. Monte Carlo simulation results show that ``long range''
plaquette decompositions are more efficient
than the ``local'' ones, especially for high levels of dissipation.
\end{abstract}

\maketitle

\section*{Introduction}
Interacting qubit models are paradigmatic examples for understanding the fundamental
physics of the quantum realm.
Since qubits are two-level quantum systems, they are simple to study while still retaining
full generality in the description of quantum phenomena.
At the same time, they also represent the basic architecture of several quantum technologies,
including quantum computation \cite{nielsen2010quantum} and quantum cryptography
\cite{RevModPhys.74.145},
which is why recent efforts have been devoted to understanding their dynamics and
thermodynamics.

On the other hand, with particular reference to quantum technologies, it should be
emphasized that such systems are never truly isolated,
but rather interact with their surrounding environment \cite{weiss2012quantum}.
Therefore, it becomes crucial to understand the effects of decoherence
and localization arising from the system–environment interaction. These effects
effectively may render the system classical,
destroying quantum coherence and compromising any effort to develop quantum technologies.
For this reason, the study of the
localization–delocalization transition induced by the environment has become of
fundamental importance, and it is typically
addressed by introducing an interaction with a bath of harmonic oscillators, which
accounts for the influence of the environment itself. In the case of the single qubit,
this gives rise to the spin-boson model
\cite{weiss2012quantum,RevModPhys.59.1,PhysRevLett.75.501,PhysRevB.58.1862}.
From the thermodynamic point of view, the spin boson is
equivalent to an Ising spin chain, with long-range interactions induced by the thermal
bath decaying as $r^{-2}$, in the case of an Ohmic bath.
When the interaction with the environment exceeds a critical value, the system undergoes
a transition of the Kosterlitz-Thouless kind \cite{PhysRevLett.37.1577}
to a magnetized state, corresponding to the localization transition of the spin boson.

An interesting question is how this behavior changes when more than one spin is coupled
to the same bosonic bath. If the spins are not directly interacting,
but the only interaction is given by the bath, then the critical value of the dissipation
scales as $N^{-1}$, where $N$ is the number of spins \cite{PhysRevB.90.224401}.
An additional ferromagnetic coupling directly between the spins further reduces the
critical value of dissipation that produces localization of the spins.
On the other hand, when $N$ spins are coupled to a common bath and interact antiferromagnetically,
the competition between the intrinsic antiferromagnetic
coupling and the bath-induced ferromagnetic interaction causes frustration
in the system \cite{DeFilippis_PRB}.
It is well-known that frustrated interactions pose a serious challenge for the study of thermodynamic properties of the system,
since Monte Carlo simulation algorithms can become highly inefficient. Cluster Monte
Carlo algorithms such as the Swendsen-Wang algorithm \cite{Swendsen_Wang_PRL}
are even more inefficient than single-spin flip algorithms in this case, due to the mismatch
between spin correlations and cluster connectivity \cite{Coniglio_PRB}.

An effective algorithm to treat frustrated models, at least fully frustrated ones,
is the Kandel \& Domany algorithm \cite{KD_fully,KD_general}. It is based
on a cluster approach, where one considers not the single pairwise interaction as
in the Swendsen-Wang algorithm, but rather a larger group of interacting spins.
In this way, one can take into account, at least partially, the frustration of the interactions.

In this paper, we investigate the efficiency of a Kandel-Domany approach to the problem
of two quantum spins, antiferromagnetically interacting with each other and coupled to the same bosonic bath.  The Hamiltonian of the system is given by
\begin{align}\label{eqn:hamiltonian}
{\cal H}=&-\frac{\Delta}{2}\left(\sigma_1^x+\sigma_2^x\right)+\frac{J}{4}\sigma_1^z\sigma_2^z
\nonumber\\
&+\left(\sigma_1^z+\sigma_2^z\right)\sum_k\lambda_k\left(a_k^\dag+a_k\right)+\sum_k\omega_ka_k^\dag
a_k,
\end{align}
where $\sigma_l^x$ and $\sigma_l^z$ with $l=1,2$ denote Pauli matrices with eigenvalues
$1$ and $-1$, and
\[
\sum_k\lambda_k^2\delta(\omega-\omega_k)=\frac{\alpha}{2}\omega_c^{1-s}\omega^s\Theta(\omega_c-\omega).
\]
We introduce the partition function of the system at temperature $T = \beta^{-1}$
\[
Z = \Tr\biggl[e^{-\beta H}\biggr],
\]
which, within the Suzuki-Trotter approximation and summing over the bosonic degrees of freedom, can be expressed as
\begin{equation}\label{eqn:partition}
Z = \sum_{\{S_{i,j}\}} e^{-{\cal S} [ \{S_{i,j}\}]}\biggl[1+o(\tau)\biggr],
\end{equation}
where $\tau=\frac{\beta}{N}$ and $N$ is the number of Trotter steps. In Eq.~\eqref{eqn:partition},
we introduced the Euclidean action of the system as
\begin{equation}\label{eqn:action}\begin{split}
&{\cal S}[\{S_{i,j}\}]=-{\tilde K}\sum_{i=0}^{N-1}\left(S_{1,i}S_{1,i+1}+S_{2,i}S_{2,i+1}\right)\\&+{\tilde
J}\sum_{i=0}^{N-1}S_{1,i}S_{2,i}
-\sum_{i<j}K_{ij}\left(S_{1,i}+S_{2,i}\right)\left(S_{1,j}+S_{2,j}\right),
\end{split}
\end{equation}
where ${\tilde K}=-\frac{1}{2}\ln\tanh\left(\frac{\tau\Delta}{2}\right)$ and ${\tilde
J}=\frac{J\tau}{4}$. In the ohmic case $s=1$, and in the limit $\omega_c\to\infty$,
$K_{ij}=\frac{\alpha\pi^2}{2N^2}\sin\left(\frac{\pi|i-j|}{N}\right)^{-2}$. We impose
periodic boundary conditions,
so that $S_{l,N}\equiv S_{l,0}$.
Thus, the system is mapped onto a frustrated double-chain Ising model up to an error
of order $\tau$. Frustration arises from the competition between the antiferromagnetic $\tilde{J}$ and ferromagnetic interactions ${\tilde K }$ and $K_{ij}$. In the
following, we set $\Delta=1$. The model described by the Hamiltonian~\eqref{eqn:hamiltonian} has been shown to undergo a quantum phase transition (QPT) in the universality class of the Berezinskii–Kosterlitz–Thouless
(BKT) transition~\cite{DeFilippis_PRB}. A QPT, which occurs at zero temperature, is
driven by quantum rather than thermal fluctuations and manifests itself through an
abrupt change in the symmetry of the system’s ground state. In the present context, at $T=0$, the model maps onto an infinite double-chain Ising system. In the absence
of dissipation, i.e. for $\alpha=0$, the system behaves as a paramagnet, since it
reduces to a one-dimensional chain with nearest-neighbor interactions only.
By increasing $\alpha$, long-range interactions are introduced, decaying as the inverse
square of the distance along the chain.
It is well established that such a decay leads to quasi-long-range correlations, characteristic
of the BKT transition
\cite{Luijten2001Criticality,Luijten2000MonteCarlo,Cardy1981OneDimModels}.
Because the two qubits share a common bath, long-range ferromagnetic interactions
$K_{ij}$ act within and across the chains. As a result, despite the
intrinsic antiferromagnetic coupling between the qubits, the two spins tend to be
aligned so that the system develops a finite magnetization for $\alpha>\alpha_c$ at
zero temperature, where $\alpha_c$ denotes the critical point of the QPT. It is important
to stress that a finite Trotter step $\tau$ only renormalizes the interaction strengths
without altering the power-law decay of $K_{ij}$. Consequently, in the discrete-time
formulation, the critical point $\alpha_c$ becomes (slightly) $\tau$ dependent, while the universality class of QPT remains unchanged.

\section*{Monte Carlo cluster algorithm method}
In this section, we will outline the Monte Carlo cluster update scheme employed in our study. A cluster algorithm generally consists of constructing clusters of spins and then flip each cluster with a given probability, usually set to $\frac{1}{2}$.
Within this framework, the efficiency of the algorithm increases when the clusters predominantly
consist of strongly correlated spins. A cluster algorithm is characterized, for any
pair of spin and bond configurations $\bigl(\{S_i\}, C\bigr)$, by a weight $W_{\text{sb}}\bigl(\{S_i\},
C\bigr)$~\cite{Cataudella_PRE}. In the following, we define $\thermal{\cdot} $ as
the conventional thermal average and $ \thermal{\cdot} _{\text{sb}}$ as the average over all spin and bond configurations, weighted by $W_{\text{sb}}(\{S_i\}, C)$. In
general, one must impose
\begin{equation}\label{eqn:cluster_mapping}
\sum_C W_{\text{sb}}(\{S_i\},C) = \exp{\biggl(-{\cal S}\bigl(\{S_i\}\bigr)\biggr)},
\end{equation}
ensuring that the average of any physical observable over the bond and spin configurations
reduces to the standard thermal average over the spin configurations. Ideally, the
algorithm should fulfill the following relations~\cite{Cataudella_PRL, Cataudella_PRE}:
\begin{equation}\label{eqn:ideal}\begin{split}
|\langle S_iS_j\rangle| = \thermal{\gamma_{ij}}_{\text{sb}}
\end{split}
\end{equation}
where we have introduced the pairwise cluster indicator $\gamma_{ij}$. For a given
configuration, $\gamma_{ij}=1$ if $i$ and $j$ belong to the same cluster, $\gamma_{ij}=0$
otherwise. By definition, $\langle \gamma_{ij}\rangle_{\text{sb}}$ represents the
connectivity between sites $i$ and $j$. In other words, Eq.~\eqref{eqn:ideal} states
that an ideal cluster algorithm exactly maps the spin-spin correlation function $|\thermal
{S_i S_j}|$ onto the percolative connectivity function  $\thermal{ \gamma_{ij}}_{\text{sb}}$.
Based on these considerations, it is straightforward to observe that as the spin system
undergoes a magnetic phase transition, the corresponding percolation model induced
by the ideal algorithm must also undergo a percolation phase transition belonging
to the same universality class and at the same critical point. In fact, this concept
underpins the entire efficiency of the update scheme. Local update algorithms, such
as the Metropolis–Hastings method, suffer from severe critical slowing down, as
the characteristic feature at criticality is the divergence of the autocorrelation
time in the thermodynamic limit. In contrast, cluster algorithms that satisfy Eq.~\eqref{eqn:ideal}
provide global and efficient updates near the critical point, thereby drastically
mitigating the effects of critical slowing down.
\subsection*{Swendsen-Wang algorithm (SW)}
In this subsection, we analyze the fundamental features of the most common cluster
algorithm, proposed by Swendsen and Wang in their pioneering work of 1987~\cite{Swendsen_Wang_PRL}.
In their work, Swendsen and Wang considered a spin system with an action of the form

\begin{equation}\label{eqn:ising}
{\cal S} \bigl(\{S_i\}\bigr) = -\sum_{\langle i j \rangle } J_{ij} S_i S_j,
\end{equation}
where in our notation, the factor $\beta$ is embedded in $J_{ij}$.
For any pair of interacting spins $i$ and $j$, a bond is introduced between them with
the following probability:
\begin{equation}\label{eqn:SW_prob}
p_{ij} =
\begin{cases}
1 - \exp(-2|J_{ij}|) & \text{if } J_{ij} S_i S_j > 0 \\
0 & \text{otherwise}.
\end{cases}
\end{equation}
In other words, the probability of forming a bond is nonzero if and only if the interaction
between the two candidate spins is satisfied, regardless of whether it is ferromagnetic
or antiferromagnetic. The algorithm based on this procedure satisfies the detailed
balance condition~\cite{Swendsen_Wang_PRL}, and thus the corresponding Markov chain
converges, after a suitable thermalization time, to the equilibrium distribution determined
by the Boltzmann weight $\exp\bigl(-{\cal S}(\{S_i\})\bigr)$. The theoretical framework
on which the algorithm is built is based on the pioneering works of Fortuin and Kasteleyn
(FK)~\cite{KF} and Coniglio and Klein (CK)~\cite{Coniglio_Klein}. These authors
demonstrated the exact mapping of a generic Ising model, as in Eq.~\eqref{eqn:ising},
onto a percolation model, where the weight of a given spin configuration $\{S_i\}$
and bond configuration $C$ is given by:
\begin{equation}\label{eqn:CK}
W_{\text{CK}}(\{S_i\},C) = \prod_{\langle ij\rangle \in C} p_{ij} \prod_{\langle lm\rangle
\in A} (1-p_{lm}),
\end{equation}
where $A$ denotes the set of all interacting spin pairs on which no bond has been
placed, and the probabilities $p_{ij}$, which depend on $S_i$ and $S_j$, are given
by Eq.~\eqref{eqn:SW_prob}. It can be shown~\cite{Coniglio_JPA, Coniglio_PRB} that
Eq.~\eqref{eqn:ideal} is exactly satisfied when using Eq.~\eqref{eqn:CK}, provided that the system is unfrustrated. This result provides the fundamental motivation for the
high efficiency of the Swendsen-Wang (SW) algorithm in unfrustrated Ising models,
as it drastically mitigates the critical slowing down associated with a possible ferromagnetic
phase transition. However, this construction fails in the presence of frustration.
Let us define the indicator $\gamma^{\parallel}_{ij}$ ($\gamma^{\not\parallel}_{ij}$),
which is equal to $1$ if the spins $i$ and $j$ belong to the same cluster and are
parallel (antiparallel) and $0$ otherwise. If the action in Eq.~\eqref{eqn:ising}
is frustrated, then, using Eq.~\eqref{eqn:CK}, one can show that:
\begin{equation}\label{eqn:frustrated}
\thermal{S_i S_j} = \thermal{\gamma^{\parallel}_{ij}}_{\text{sb}} - \thermal{\gamma^{\not\parallel}_{ij}}_{\text{sb}},
\end{equation}
which, together with the identity $\gamma_{ij} = \gamma^{\parallel}_{ij} + \gamma^{\not\parallel}_{ij}$,
leads to:
\begin{equation}\label{eqn:frustrated_ineq}
|\thermal{S_i S_j}| \leq \thermal{\gamma_{ij}}_{\text{sb}}.
\end{equation}
As specified before, Eq.~\eqref{eqn:frustrated_ineq} holds as an inequality only in
the presence of frustration. In its absence, a given pair of spins at sites $i$ and
$j$ can be connected only if they are parallel (or antiparallel, depending on the
lattice geometry), regardless of the specific bond path connecting them. Thus, only
one between $\gamma_{ij}^{\parallel}$ and $\gamma_{ij}^{\not\parallel}$ contributes
to Eq.~\eqref{eqn:frustrated}, turning Eq.~\eqref{eqn:frustrated_ineq} into an equality.
On the other hand, frustration allows for configurations where two antiparallel spins
at sites $i$ and $j$ are connected by any number of ferromagnetic bonds and an odd
number of antiferromagnetic bonds, giving a non-zero contribution to $\gamma^{\not\parallel}_{ij}$;
similarly, the same spins, if parallel, can be connected by any number of ferromagnetic
bonds but by an even number of antiferromagnetic bonds, resulting in a non-zero contribution
to $\gamma^{\parallel}_{ij}$. Both types of configurations contribute positively to
the connectivity $\thermal{\gamma_{ij}}_{\text{sb}}$, but give opposite contributions
to $\thermal{S_i S_j}$, thereby clarifying the geometrical meaning of Eq.~\eqref{eqn:frustrated_ineq}.

\subsection*{Kandel-Domany algorithm and plaquette decomposition}
Kandel and Domany were the first to develop an algorithm capable of efficiently simulating
the fully frustrated Ising model defined on a two-dimensional square lattice. This
model is described by an action of the form \eqref{eqn:ising}, where $|J_{ij}| = J
> 0$ for all nearest-neighbor pairs $i$ and $j$. The coupling sign is always positive
for horizontal interactions, while for vertical interactions it is positive on even
columns and negative on odd columns. In this case, frustration is present in each
of the smallest square plaquettes $q$ of the lattice, since $\prod_{\thermal{ij} \in
q} \mathrm{sgn}(J_{ij}) = -1$. Unlike the SW algorithm, the criterion for bond assignment
on the lattice is not based on considering interacting spin pairs, but rather on partitioning
the lattice into a checkerboard pattern of black and white plaquettes. Once either
the black or the white plaquettes are chosen, for each plaquette one inserts, with
a probability depending on the spin configuration, either no bond, a single bond,
or two bonds~\cite{KD_fully}. The result is a procedure that reduces the lattice connectivity,
as it excludes the possibility of connecting all spins within a single plaquette.
Subsequent works by the same authors~\cite{KD_general} and by others~\cite{Cataudella_PRE}
have shown that the algorithm proposed by Kandel and Domany is a special case of a
broader class of algorithms that generalize the simpler SW scheme. The main idea behind
these algorithms is to rewrite the action \eqref{eqn:ising} as a sum over contributions
of elementary units $l$: 
\begin{equation}\label{eqn:plaq_decomp}
{\cal S} [\{S_i\}] = \sum_l {\cal S}^l [\{S_i^l\}].
\end{equation}
In general, $l$ may represent a single interacting spin pair, a plaquette containing
multiple interacting spins, \textit{et cetera}. Moreover, the same interaction between
a given pair $\thermal{ij}$ may be subdivided by multiple elementary units, with the
only requirement that the sum of all such contributions reproduces the original interaction,
i.e., $\sum_l J_{ij}^l = J_{ij}$. For each elementary unit $l$, one considers all
possible bond graphs $\eta^l$, assigning to each a weight $w_{\eta^l} > 0$ independent
of the spin configuration of that unit. A graph $\eta^l$ is said to be \textit{compatible}
with a spin configuration $\{S_i^l\}$ if no bonds are placed between spins whose interaction
is unsatisfied; otherwise, it is \textit{incompatible}. In general, the weights must
satisfy the following equation
\begin{equation}\label{eqn:weights}
\sum_{\eta^l} w_{\eta^l} \, \delta_{\{S_i^l\} \eta^l} = \exp\biggl(-{\cal S}^l \bigl[\{S_i^l\}\bigr]\biggr),
\end{equation}
where $\delta_{\{S_i^l\} \eta^l} = 1$ if $\{S_i^l\}$ and $\eta^l$ are compatible,
and $0$ otherwise. The conditional probability of assigning graph $\eta^l$ given the
spin configuration $\{S_i^l\}$ is then defined as
\begin{equation}\label{eqn:KD_prob}
P\bigl(\eta^l | \{S_i^l\}\bigr) = \frac{w_{\eta^l} \, \delta_{\{S_i^l\} \eta^l}}{\exp\biggl(-{\cal
S}^l \bigl[\{S_i^l\}\bigr]\biggr)},
\end{equation}
which is evidently normalized and lies between $0$ and $1$ by virtue of Eq.~\eqref{eqn:weights}.
We note, moreover, that Eq.~\eqref{eqn:weights} is a local version of Eq.~\eqref{eqn:cluster_mapping}.
In each Monte Carlo step, given the initial spin configuration $\{S_i\}$, every elementary
unit $l$ is examined and a bond graph $\eta^l$ is assigned with the probability given
by Eq.~\eqref{eqn:KD_prob}, thereby contributing to the bond configuration of the
entire lattice. It is straightforward to verify that, when the elementary unit consists
of a single pair of interacting spins, Eq.~\eqref{eqn:SW_prob} is the unique solution
of Eq.~\eqref{eqn:weights} and Eq.~\eqref{eqn:KD_prob}. Furthermore, it has been shown~\cite{Cataudella_PRE}
that Eq.~\eqref{eqn:weights}, together with Eq.~\eqref{eqn:KD_prob}, satisfies Eq.~\eqref{eqn:cluster_mapping}
and ensures that the algorithm fulfills the detailed balance condition~\cite{KD_general}.
When sufficiently large elementary cells are considered, Eq.~\eqref{eqn:weights} may
no longer be sufficient to uniquely determine the weights $w_{\eta^l}$. In such cases,
one can choose the solution that satisfies Eq.~\eqref{eqn:ideal}, if it exists, or,
at least, the one that minimizes the distance between the right-hand side and the
left-hand side of Eq.~\eqref{eqn:frustrated_ineq}.
\section*{Application of plaquette decomposition}
As previously discussed, our model~\eqref{eqn:hamiltonian} is equivalent to a frustrated
double-chain Ising model~\eqref{eqn:action}. Consequently, conventional cluster algorithms
based on SW update schemes fail at criticality. In order to reduce the slowing down
as much as possible, we employed a generalized Kandel-Domany approach. The plaquette
partitioning introduced in Eq.~\eqref{eqn:plaq_decomp} and the subsequent resolution
of Eq.~\eqref{eqn:weights} have been addressed by proposing several criteria and approaches,
which will be discussed later in the following subsections. An analytical attempt
to solve Eq.~\eqref{eqn:weights} was made by partitioning the double-chain into triangular
and rectangular plaquettes, enforcing vanishing weights for graphs containing bonds
between antiferromagnetically interacting spins and, whenever possible, for graphs
with high connectivity. This choice is justified by the fact that the system undergoes
a quantum phase transition into a magnetized phase. Hence, close to the transition,
spins are more likely to be correlated rather than anticorrelated. Ideally, one should
devise a method to exactly suppress any possibility of connecting antiparallel spins,
thereby turning Eq.~\eqref{eqn:frustrated_ineq} into Eq.~\eqref{eqn:ideal}. Since
we choose partitionings which do not, in principle, fully satisfy the original interactions,
if necessary, we also introduced SW-like elementary units, each consisting of a pair
of interacting spins, in order to complete the interactions. In the first approach,
we consider ``long-range" triangular or rectangular plaquettes. In the triangular
case, the plaquette consists of two nearest neighboring spins on opposite sides of
the double chain (``opposing" spins), and one spin at long distance on one of the
two chains (see Fig.~\ref{fig:triangular_plaq}). In the rectangular case, we consider
two pairs of opposing spins at long distance (see Fig.~\ref{fig:rectangular_plaq}).
In both cases, each pair of opposing spins belongs to a large number of different
plaquettes, having different ``heights" along the chain. Thus, a variable number of
elementary plaquettes subdivide the same antiferromagnetic interaction, depending
on temperature and on the tunable Hamiltonian parameters. Since this method fully
(or at least largely) accounts for the long-range interaction, we will hereafter refer
to it as the ``long-range'' plaquette decomposition. In the second approach, we employ
nearest-neighbor rectangular plaquettes (see Fig.~\ref{fig:rectangular_plaq}) and,
alternatively, a bi-rectangular unit consisting of two consecutive nearest-neighbor
rectangular plaquettes (see Fig.~\ref{fig:birectangular_plaq}). In the first case,
each antiferromagnetic interaction is subdivided equally by two adjacent plaquettes;
in the second case, each antiferromagnetic interaction is subdivided equally by three
consecutive bi-rectangular units, while any ferromagnetic interaction of the unit
is subdivided equally by two consecutive units. In both cases, SW-like elementary
units were introduced to account for any long-range interaction not covered by the
rectangular or bi-rectangular plaquettes. Since this method accounts only for short-range
contributions of the kernel $K_{ij}$, we will hereafter refer to it as the ``local''
plaquette decomposition. For the long-range approach, we solved Eq.~\eqref{eqn:weights}
analytically, by assigning zero weight to graphs containing antiferromagnetic bonds
and, if possible, suppressing those with excessive connectivity. In contrast, in the
local approach, Eq.~\eqref{eqn:weights} was solved numerically by employing a linear
programming technique, which can be used to optimize a linear function subject to
constraints that define a convex feasible region. In this way, we identified one or
more linear functions that reduce the distance between the right-hand side and the
left-hand side of Eq.~\eqref{eqn:frustrated_ineq}, subject to the equations for $w_\eta^l$.

\subsection*{Long-range decomposition: triangular plaquettes (LR3)}
\begin{figure}
\centering
\includegraphics[width=\columnwidth]{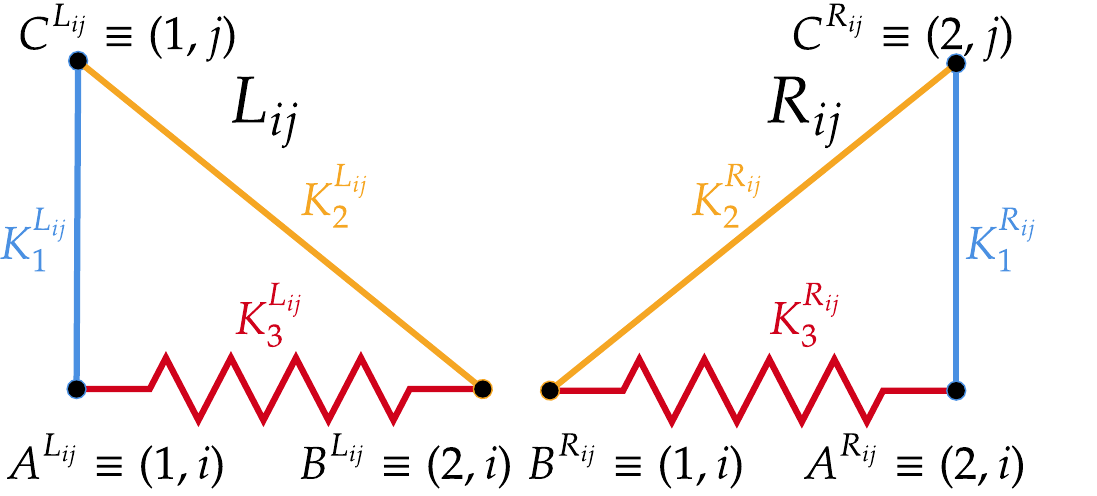}
\caption{Geometry of the triangular plaquettes $L_{ij}$ (left) and $R_{ij}$ (right),
employed for LR3. Straight lines represent ferromagnetic interactions, while zig-zag
ones stand for antiferromagnetic interactions. It is worth noting that site $j$ may
lie arbitrarily far from site $i$.}
\label{fig:triangular_plaq}
\end{figure}
In this approach, we partition the lattice using triangular plaquettes formed by the
sets of sites $L_{ij}=\{(1, i), (2, i), (1, j)\}$ and $R_{ij}=\{(1, i), (2, i), (2,
j)\}$. It should be noted that, due to periodic boundary conditions, in order to partition
the entire lattice using only triangular plaquettes, it is sufficient to consider,
for each possible value of $i$ from $0$ to $N-1$, plaquettes such that $j=\mod_N(i+k)$
with $k=1,\dots,\lfloor N/2\rfloor$, assuming $N$ is odd. In what follows, we adopt
the site and interaction nomenclature introduced in Fig.~\ref{fig:triangular_plaq}.
The action of such plaquettes reads:
\begin{equation}\begin{split}\label{eqn:triangular_action}
&{\cal S}^{L_{ij}}[\{S^{L_{ij}}_{lm}\}] =\\ 
& - K^{L_{ij}}_1 S_{A^{L_{ij}}}S_{C^{L_{ij}}} - K^{L_{ij}}_2 S_{B^{L_{ij}}}S_{C^{L_{ij}}}
+ K^{L_{ij}}_3 S_{A^{L_{ij}}}S_{B^{L_{ij}}} \\
&{\cal S}^{R_{ij}}[\{S^{R_{ij}}_{lm}\}] =\\
&- K^{R_{ij}}_1 S_{A^{R_{ij}}}S_{C^{R_{ij}}} - K^{R_{ij}}_2 S_{B^{R_{ij}}}S_{C^{R_{ij}}}
+ K^{R_{ij}}_3 S_{A^{R_{ij}}}S_{B^{R_{ij}}},
\end{split}
\end{equation}
\begin{figure}
\centering
\includegraphics[width=\columnwidth]{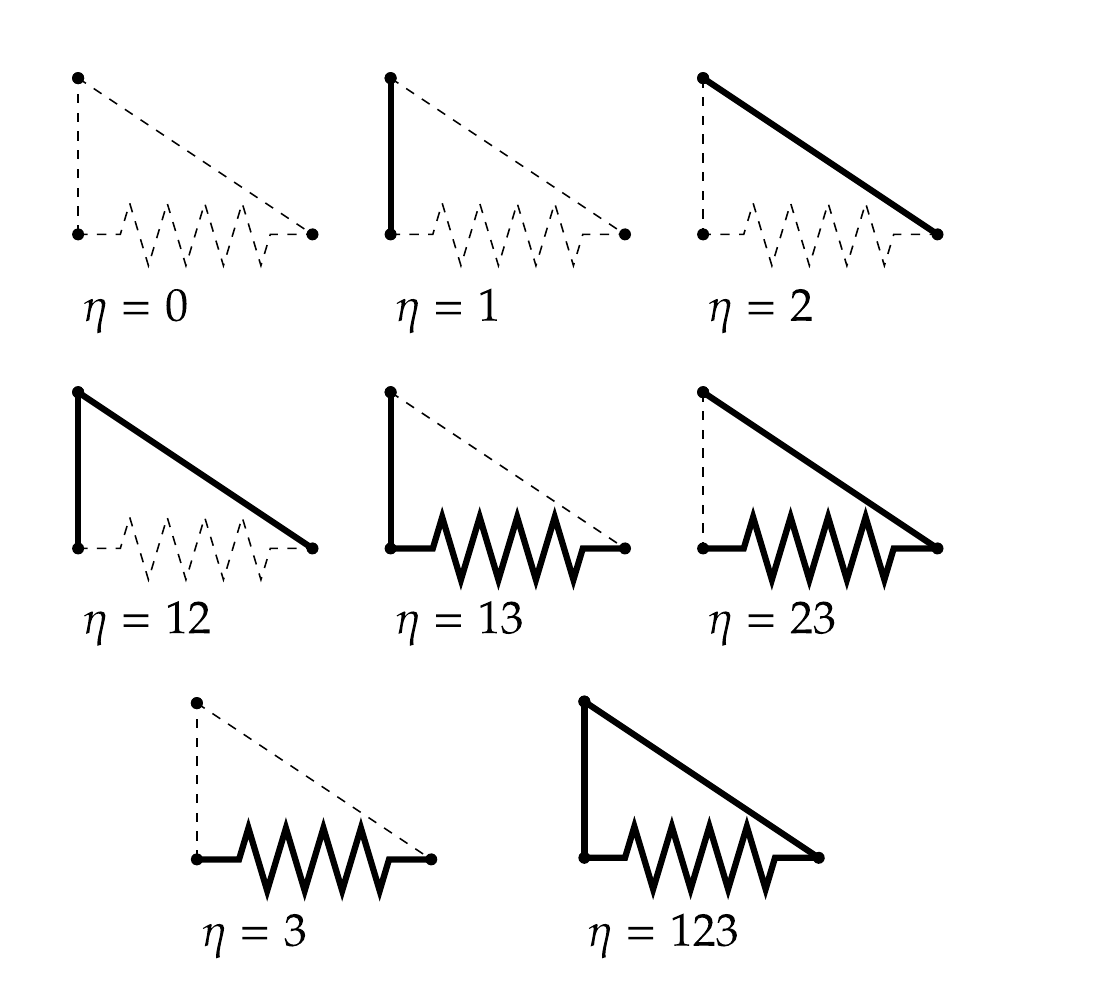}
\caption{All possible graphs $\eta$ for $L_{ij}$. Dashed lines indicate missing bonds,
while thick solid lines denote present bonds. Similarly, we define, by symmetry, analogous
graphs for $R_{ij}$ with the same labels.}
\label{fig:triangular_graph}
\end{figure}
where we define $K^{T_{ij}}_1 \equiv K^{L_{ij}}_1=K^{R_{ij}}_1 = K_{ij} + \tilde{K}\delta_{j\,i+1}$,
and $K^{T_{ij}}_2 \equiv K^{L_{ij}}_2=K^{R_{ij}}_2 = K_{ij}$. Moreover, we also define
$K^{T_{ij}}_3 = K^{L_{ij}}_3 = K^{R_{ij}}_3$, whose optimal value will be determined
later. For each plaquette $L_{ij}$ and $R_{ij}$ we assign a weight to the graphs introduced
in Fig.~\ref{fig:triangular_graph}. Since the interactions of $L_{ij}$ and $R_{ij}$
are equivalent, we assign the same weights to the equivalent graphs of both plaquettes,
i.e., $w^{T_{ij}}_\eta \equiv w^{L_{ij}}_\eta = w^{R_{ij}}_\eta$. Thus, Eq.~\eqref{eqn:weights}
becomes:
\begin{equation}\begin{split}\label{eqn:triangular_weights}
&w^{T_{ij}}_0 = \exp{\biggl(-K^{T_{ij}}_1 - K^{T_{ij}}_2 - K^{T_{ij}}_3\biggr)} \\
&w^{T_{ij}}_0 + w^{T_{ij}}_1 + w^{T_{ij}}_2 + w^{T_{ij}}_{12} = \exp{\biggl(K^{T_{ij}}_1
+ K^{T_{ij}}_2 - K^{T_{ij}}_3\biggr)} \\
&w^{T_{ij}}_0 + w^{T_{ij}}_1 + w^{T_{ij}}_3 + w^{T_{ij}}_{13} = \exp{\biggl(K^{T_{ij}}_1
- K^{T_{ij}}_2 + K^{T_{ij}}_3\biggr)} \\
&w^{T_{ij}}_0 + w^{T_{ij}}_2 + w^{T_{ij}}_3 + w^{T_{ij}}_{23} = \exp{\biggl(-K^{T_{ij}}_1
+ K^{T_{ij}}_2 + K^{T_{ij}}_3\biggr)}.
\end{split}
\end{equation}
Note that $w^{T_{ij}}_{123}$ is incompatible with any spin configuration, as expected,
since the chosen plaquette is frustrated. Therefore, it is assigned a vanishing weight.
To minimize connectivity, i.e. the right-hand side of~\eqref{eqn:frustrated_ineq},
we impose that graphs containing at least one bond on the antiferromagnetic interaction
are set to zero, i.e., $w^{T_{ij}}_3 = w^{T_{ij}}_{13} = w^{T_{ij}}_{23} = 0$. Moreover,
to further reduce it, we assign to the plaquette an interaction $K^{T_{ij}}_3 = \bar{K}^{T_{ij}}_3$
such that the weight $w^{T_{ij}}_{12}$ vanishes. This is justified since the corresponding
graph connects all spins in the plaquette, thus maximizing the connectivity. This
choice yields the following solution:
\begin{equation}\begin{split}\label{eqn:triangular_solution}
&w^{T_{ij}}_0 = \exp{\biggl(-K^{T_{ij}}_1 - K^{T_{ij}}_2 - K^{T_{ij}}_3\biggr)} \\
&w^{T_{ij}}_1 = 2e^{-K^{T_{ij}}_2}\sinh\biggl(K^{T_{ij}}_1 + \bar{K}^{T_{ij}}_3\biggr)
\\
&w^{T_{ij}}_2 = 2e^{-K^{T_{ij}}_1}\sinh\biggl(K^{T_{ij}}_2 + \bar{K}^{T_{ij}}_3\biggr)
\\
&\bar{K}^{T_{ij}}_3 = \frac{1}{2}\ln\biggl\{\frac{\cosh\bigl(K^{T_{ij}}_1+K^{T_{ij}}_2\bigr)}{\cosh\bigl(K^{T_{ij}}_1-K^{T_{ij}}_2\bigr)}\biggr\}.
\end{split}
\end{equation}
It is worth emphasizing that the equations must be solved for all values of $j$, but
only for a single value of $i$, as the system is translationally invariant along the
chain. Once the weights are determined, we can choose to partition the entire lattice
with triangular plaquettes, assigning the bonds to the lattice considering $L_{ij}$
and $R_{ij}$ with $i=0,\dots,N-1$ and $j=\mod_N\bigl(i+1\bigr),\dots,\mod_N\bigl ({i+\lfloor
N/2\rfloor} \bigr)$. To satisfy Eq.~\eqref{eqn:plaq_decomp}, one must, for each $i$,
introduce a new SW-like elementary unit between spins $S_{1,i}$ and $S_{2,i}$ of action:
\begin{equation}\label{eqn:triangular_res_0}
{\cal S}^{T,\text{res}}_i[S_{1,i}, S_{2,i}] = K^{T,\text{res}} S_{1,i} S_{2,i},
\end{equation}
where $K^{T, \text{res}} = \tilde{J} - 2\sum_{j=1}^{\lfloor N/2\rfloor} \bar{K}_3^{T_{0j}}$
is the residual interaction. We stress that the residual interaction may in some cases
become negative, effectively rendering the interaction in~\eqref{eqn:triangular_res_0}
ferromagnetic. An alternative approach consists of identifying the maximum value $j_\text{max}$
such that $K^{T, \text{res}} = \tilde{J} > 2\sum_{j=1}^{j_\text{max}} \bar{K}^{T_{0j}}_3$.
In this case, the lattice is covered with $2j_\text{max}$ triangular plaquettes for
each $i$, and a new SW-like elementary unit is introduced between spins $S_{1,i}$
and $S_{2,i}$, with action:
\begin{equation}\label{eqn:triangular_res_1}
{\cal S}^{T,\text{res}}_i[S_{1,i}, S_{2,i}] = \hat{K}^{T,\text{res}} S_{1,i} S_{2,i},
\end{equation}
where $\hat{K}^{T,\text{res}} = \tilde{J} - 2\sum_{j=1}^{j_\text{max}} \bar{K}_3^{T_{0j}}$.

Finally, for all remaining long-range interactions not explicitly included, we resort
to SW-like elementary units. In a subsequent performance analysis of the algorithms,
we will consider both the case in which the residual interaction may change sign,
and the one in which it is constrained to remain antiferromagnetic.
\subsection*{Long-range decomposition: rectangular plaquettes (LR4)}
\begin{figure}
\centering
\includegraphics[width=\columnwidth]{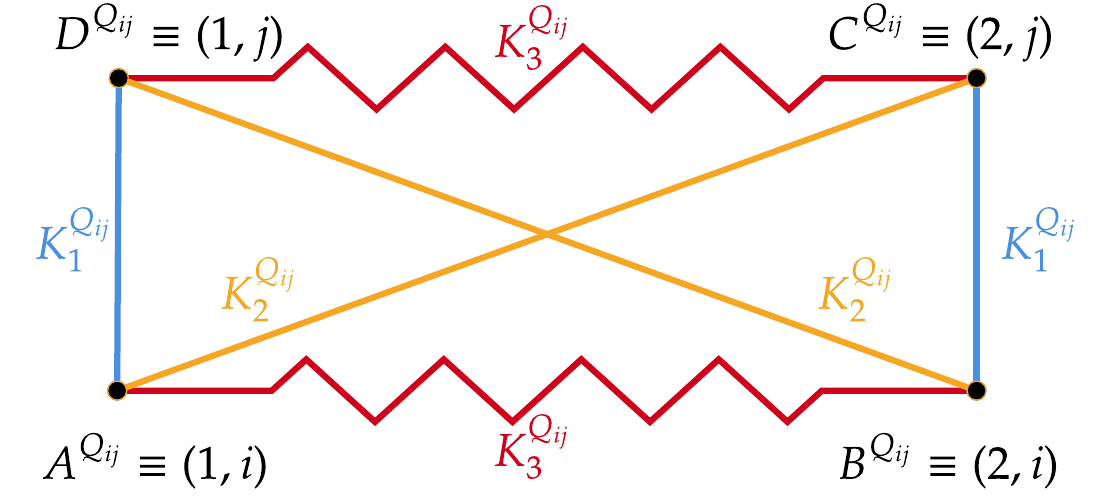}
\caption{Geometry of the rectangular plaquette $Q_{ij}$. Straight lines represent
ferromagnetic interactions, while zig-zag ones stand for antiferromagnetic interactions.
Same color are employed for interactions which always have same magnitude. This type
of plaquette is employed for both LR4, if site $j$ can lie arbitrarily far from site
$i$, and LOC4, if site $j=i+1$, taken modulo $N$.}
\label{fig:rectangular_plaq}
\end{figure}
In this approach, we partition the lattice using rectangular plaquettes formed by
the set of sites $Q_{ij}=\{(1, i), (2, i), (2, j),(1,j)\}$. Using the same reasoning
as previously adopted for the triangular ones, it is sufficient to consider $N\lfloor
N/2 \rfloor$ rectangular plaquettes to partition the entire lattice. In the following,
we adopt the site and interaction nomenclature of such a plaquette introduced in Fig.~\ref{fig:rectangular_plaq}.
Its action reads:
\begin{equation}\begin{split}\label{eqn:rectangular_action}
&{\cal S}^{Q_{ij}}[\{S^{Q_{ij}}_{lm}\}] =- K^{Q_{ij}}_1 (S_{A^{Q_{ij}}}S_{D^{Q_{ij}}}+S_{B^{Q_{ij}}}S_{C^{Q_{ij}}})\\
&- K^{Q_{ij}}_2 (S_{B^{Q_{ij}}}S_{D^{Q_{ij}}} + S_{A^{Q_{ij}}}S_{C^{Q_{ij}}})\\
&+K^{Q_{ij}}_3 (S_{A^{Q_{ij}}}S_{B^{Q_{ij}}}+S_{C^{Q_{ij}}}S_{D^{Q_{ij}}}).
\end{split}
\end{equation}
\begin{figure}
\centering
\includegraphics[width=\columnwidth]{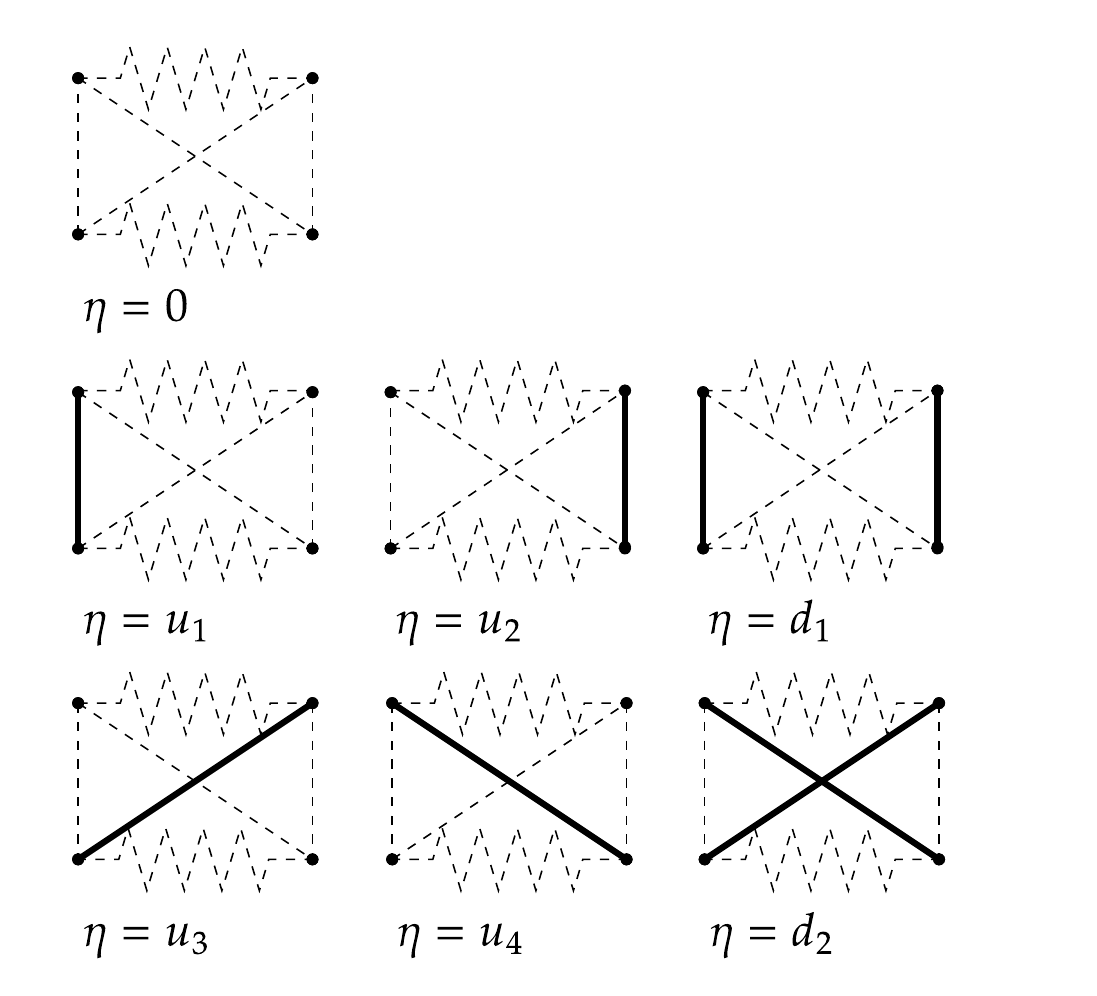}
\caption{All graphs $\eta$ with non vanishing weights considered for $Q_{ij}$ in the
LR4 approach. Dashed lines indicate missing bonds, while thick solid lines denote
present bonds.}
\label{fig:rectangular_graph}
\end{figure}
We set $K^{Q_{ij}}_f
\equiv K^{Q_{ij}}_1 = K^{Q_{ij}}_2 = K_{ij}$, leaving out the short-range interaction
$\tilde{K}$, which was treated separately using SW-like elementary units. For each
plaquette, we set $K^{Q_{ij}}_3 = \bar{K}^{Q_{ij}}_3$, chosen to enforce a vanishing
weight for graphs containing bonds on antiferromagnetic interactions, as well as for
certain highly connected graphs. Specifically, we excluded the graph in which bonds
occupy all four ferromagnetic interactions simultaneously, and the graphs with two
bonds such that one lies on the same chain while the other connects the two chains.
On a single plaquette, these configurations generate, respectively, a four-site cluster
and a three-site cluster. The only graphs considered are those shown in Fig.~\ref{fig:rectangular_graph},
which, within a single unit, generate clusters of at most two sites. Although there
are $7$ possible graphs, the number of associated unknowns is only $3$. Indeed, since
we chose $K^{Q_{ij}}_1 = K^{Q_{ij}}_2$, symmetry under site exchange implies that
graphs $u_1$, $u_2$, $u_3$, and $u_4$ are equivalent, as are $d_1$ and $d_2$. Therefore,
we define $w_u^{Q_{ij}} = w_{u_1}^{Q_{ij}} = w_{u_2}^{Q_{ij}} = w_{u_3}^{Q_{ij}} =
w_{u_4}^{Q_{ij}}$ and $w_d^{Q_{ij}} = w_{d_1}^{Q_{ij}} = w_{d_2}^{Q_{ij}}$. Thus,
Eq.~\eqref{eqn:weights} becomes:
\begin{equation}\begin{split}\label{eqn:rectangular_weights}
&w^{Q_{ij}}_0 + 4w^{Q_{ij}}_u +2w^{Q_{ij}}_d = \exp{\biggl(4K^{Q_{ij}}_f -  2\bar{K}^{Q_{ij}}_3\biggr)}
\\
&w^{Q_{ij}}_0 + 2w^{Q_{ij}}_u  = 1 \\
&w^{Q_{ij}}_0 + 2w^{Q_{ij}}_u + w^{Q_{ij}}_d = \exp{\biggl(2\bar{K}^{Q_{ij}}_3\biggr)}
\\
&w^{Q_{ij}}_0 = \exp{\biggl(-4K^{Q_{ij}}_f -2\bar{K}^{Q_{ij}}_3\biggr)},
\end{split}
\end{equation}
whose solution is given by
\begin{equation}\begin{split}\label{eqn:rectangular_solution}
&w^{Q_{ij}}_0 = \exp{\biggl(-4K^{Q_{ij}}_f -2\bar{K}^{Q_{ij}}_3\biggr)} \\
&w^{Q_{ij}}_u = \frac{1}{2}-\frac{1}{2}\exp{\biggl(-4K^{Q_{ij}}_f -2\bar{K}^{Q_{ij}}_3\biggr)}
\\
&w^{Q_{ij}}_d = \exp{\biggl(2\bar{K}^{Q_{ij}}_3\biggr)} -1\\
&\bar{K}^{Q_{ij}}_3 = \frac{1}{4}\ln\biggl\{\cosh{\biggl( 4K_f^{Q_{ij}}\biggr)}\biggr\}.
\end{split}
\end{equation}
As in the case of triangular plaquettes, we proceed with two different approaches.
The first consists of tiling the entire lattice with $N \lfloor N/2 \rfloor$ rectangular
plaquettes, and introducing SW-like units to account for the $\tilde{K}$ short-range
interaction and for the residual interaction $K^{Q,\text{res}} = \tilde{J} - 2\sum_{j=1}^{\lfloor
N/2 \rfloor} \bar{K}_3^{Q_{0j}}$. The second approach again considers the maximum
integer $j_{\text{max}}$ such that the residual interaction $\hat{K}^{Q,\text{res}}
= \tilde{J} - 2\sum_{j=1}^{j_{\text{max}}} \bar{K}_3^{Q_{0j}}$ remains positive. As
before, the remaining long-range interactions are covered by the SW procedure. 
\subsection*{Local decomposition: rectangular plaquettes (LOC4)}
In this approach, we partition the lattice into rectangular plaquettes $Q_{i\,i+1}$
for $i = 0, \dots, N-1$, and $i+1$ taken modulo $N$, with the remaining interactions
treated using the SW algorithm. To satisfy Eq.~\eqref{eqn:plaq_decomp}, we set $K^{Q_{i\,i+1}}_3
= \tilde{J}/2$, $K^{Q_{i\,i+1}}_1 = K_{i\,i+1} + \tilde{K}$, and $K^{Q_{i\,i+1}}_2
= K_{i\,i+1}$. In this arrangement, every ferromagnetic interaction is equally subdivided
between two plaquettes. It is worth noting that choosing $K^{Q_{i\,i+1}}_1 \neq K^{Q_{i\,i+1}}_2$
 breaks some of the symmetries previously exploited in the LR4 approach. In
this framework, we do not set to zero any graph weight a priori.
Instead, we solve Eq.~\eqref{eqn:weights} numerically using the simplex algorithm
\cite{press_etal},
minimizing a selected cost function (introduced below) subject to the constraints
of the equation. Specifically, we define two cost functions:
\begin{equation}\begin{split}\label{eqn:cost_functions}
&h_1(\eta) = B_A(\eta), \\
&h_2(\eta) = \sum_{s=1}^{N_\text{plaq}} \frac{N_s(\eta) s^2}{N_\text{plaq}^2},
\end{split}
\end{equation}
where $B_A(\eta)$ denotes the number of bonds in graph $\eta$ lying on antiferromagnetic
interactions, $N_s(\eta)$ the number of clusters of size $s$ in graph $\eta$ and the
$N_\text{plaq}$ the total number of sites of the plaquette. The use of $h_1(\eta)$
allows us to select graphs that exclude direct connections between spins coupled antiferromagnetically,
thereby suppressing contributions to $\gamma_{ij}^{\not \parallel}$. In fact, completely
suppressing any possibility of connecting antiparallel spins, turns Eq.~\eqref{eqn:frustrated_ineq}
into Eq.~\eqref{eqn:ideal}, as already specified in the previous sections. On the
other hand, $h_2(\eta)$ directly penalizes highly connected graphs, favoring configurations
with low connectivity. The choice of the weights may depend on the physical parameters
and on the temperature.
\subsection*{Local decomposition: bi-rectangular plaquettes (LOC6)}
\begin{figure}
\centering
\includegraphics[width=\columnwidth]{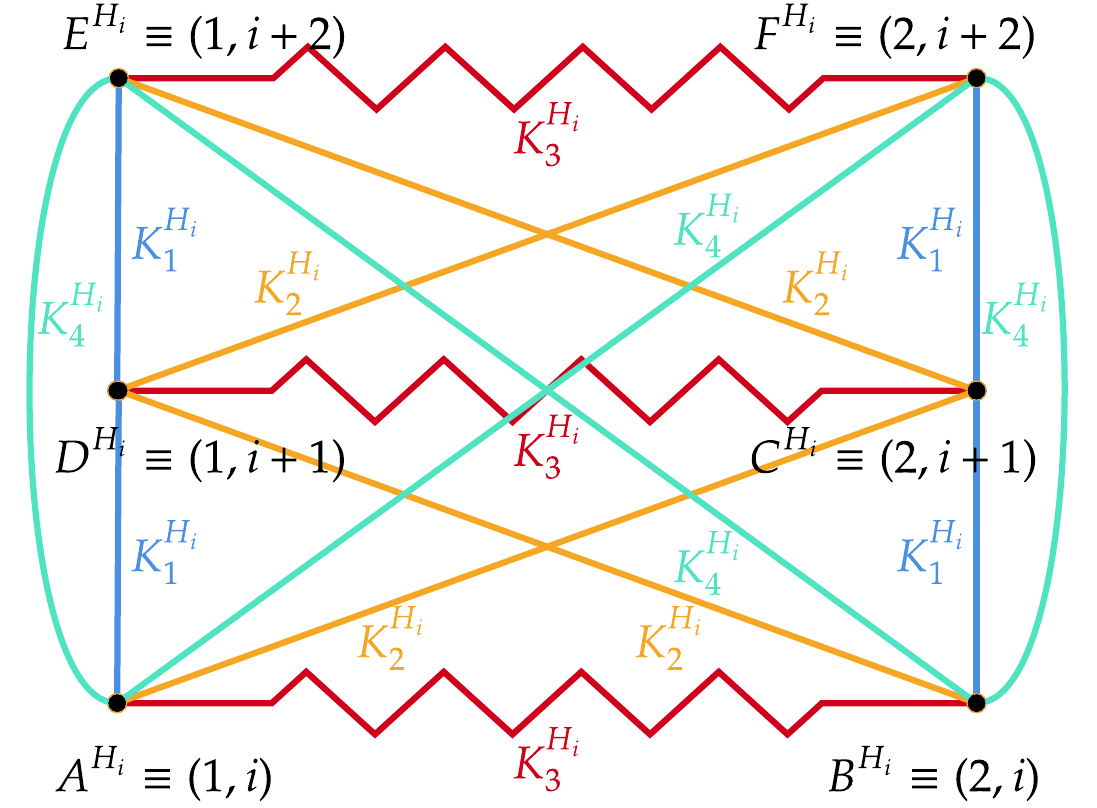}
\caption{Geometry of the bi-rectangular plaquette $H_{i}$, employed for LOC6. Straight
lines represent ferromagnetic interactions, while zig-zag ones stand for antiferromagnetic
interactions. Same color are employed for interactions which always have same magnitude.}
\label{fig:birectangular_plaq}
\end{figure}
In this approach, we partition the lattice using bi-rectangular plaquettes defined
by the set of sites $H_{i}=\{(1, i), (2, i), (2, i+1), (1,i+1), (1,i+2), (2,i+2)\}$,
where the indices $i+1$ and $i+2$ are taken modulo $N$. The nomenclature adopted here
is illustrated in Fig.~\ref{fig:birectangular_plaq}. The action associated to the
plaquette reads:
\begin{equation}\begin{split}\label{eqn:bi_rectangular_action}
&{\cal S}^{H_i}[\{S^{H_i}_{lm}\}] =\\
&- K^{H_i}_1 (S_{A^{H_i}}S_{D^{H_i}}+S_{B^{H_i}}S_{C^{H_i}}+S_{C^{H_i}}S_{F^{H_i}}+S_{D^{H_i}}S_{E^{H_i}})\\
& - K^{H_i}_2 (S_{A^{H_i}}S_{C^{H_i}}+S_{B^{H_i}}S_{D^{H_i}}+S_{C^{H_i}}S_{E^{H_i}}+S_{D^{H_i}}S_{F^{H_i}})\\
&+K^{H_i}_3 (S_{A^{H_i}}S_{B^{H_i}}+S_{C^{H_i}}S_{D^{H_i}}+S_{E^{H_i}}S_{F^{H_i}})\\
& - K^{H_i}_4 (S_{A^{H_i}}S_{F^{H_i}}+S_{B^{H_i}}S_{E^{H_i}}+S_{A^{H_i}}S_{E^{H_i}}+S_{B^{H_i}}S_{F^{H_i}}).
\end{split}
\end{equation}

\begin{figure}[t!]
\centering
\includegraphics[]{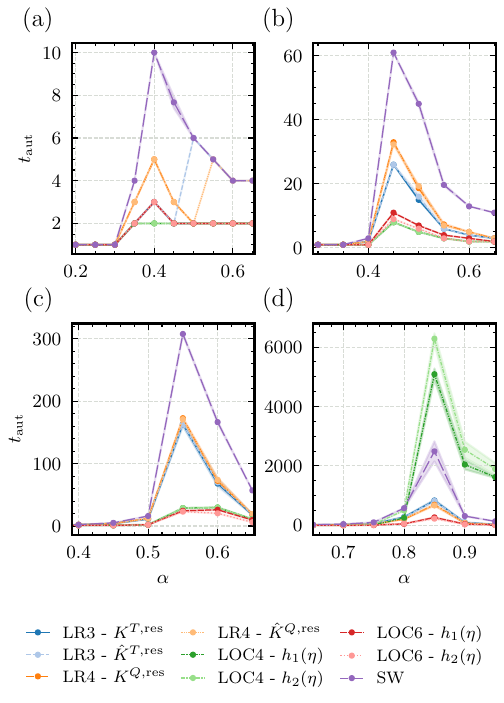}
\caption{Autocorrelation time $t_{\text{aut}}$ as a function of $\alpha$ for $J=1$
(a), $J=2$ (b), $J=3$ (c), $J=5$ (d) for the partitioning of the lattice introduced
in the main text. Shaded regions around each point represent the statistical error
computed on three independent simulations.}
\label{fig:aut_time}
\end{figure}

Since bi-rectangular plaquettes involve $6$ sites and $15$ interactions, an analytical
treatment of Eq.~\eqref{eqn:weights} is computationally prohibitive. Therefore, we
employ a numerical approach analogous to that used in the LR4 approach. The
lattice is tiled by plaquettes $H_i$ for $i = 0, \dots, N-1$, where $K^{H_i}_3 = \tilde{J}/3$,
$K^{H_i}_1 = (K_{i\,i+1} + \tilde{K})/2$, $K^{H_i}_2 = K_{i\,i+1}/2$, $K^{H_i}_4 =
K_{i\,i+2}$. In this way, each interaction is equally subdivided among the plaquettes
that include it. As for rectangular plaquettes, Eq.~\eqref{eqn:weights} is solved
twice using the simplex algorithm, minimizing the same cost functions defined in Eq.~\eqref{eqn:cost_functions}.
Any remaining long-range interactions are again treated via the SW algorithm.
\begin{figure}[t!]
\centering
\includegraphics[]{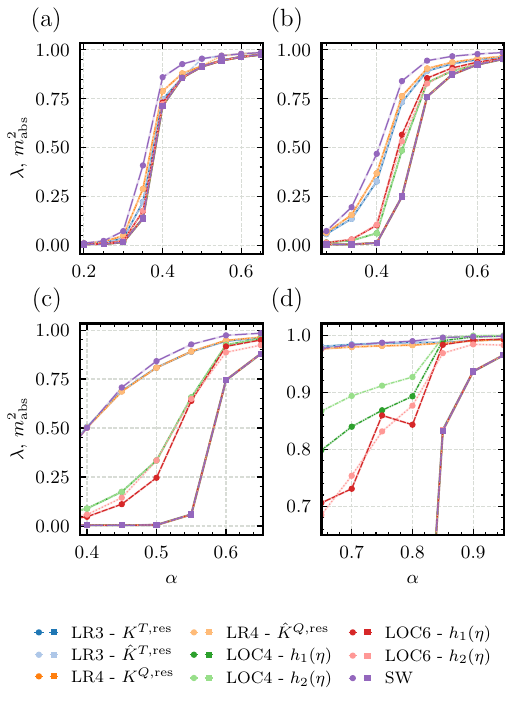}
\caption{Comparison between $\lambda$ (circles) and $m^2_{\text{abs}}$ (squares) as
functions of $\alpha$ for $J=1$ (a), $J=2$ (b), $J=3$ (c), $J=5$ (d) for the partitioning
of the lattice introduced in the main text. Shaded regions around each point represent
the statistical error computed on three independent simulations.}
\label{fig:mcs}
\end{figure}
\section*{Results}
In this section, we analyze the results of each clustering technique introduced in
the previous section. Specifically, we consider the autocorrelation time $t_{\text{aut}}$
as the main indicator of algorithmic efficiency. To define it, we first introduce
the autocorrelation function under analysis:
\begin{equation}\label{eqn:aut_func}
C(t-t') = \frac{\thermal{|m(t)||m(t')|} - \thermal{|m|}^2}{\thermal{|m|^2} - \thermal{|m|}^2}
\end{equation}
where $|m(u)|$ denotes the observable $|m|$ computed at the $u$-th Monte Carlo step.
In particular, we define $|m| = \frac{1}{2N} \left| \sum_i S_i \right|$, where $2N$
is the total number of lattice sites. To compute this function efficiently, we exploit
the time-translation invariance of Monte Carlo simulations, using multiple starting
steps $t'$. Since $|m|$ is related to the order parameter of the quantum phase transition,
we expect $t_{\text{aut}}$ to exhibit a maximum at criticality for finite $N$, i.e.,
finite $\beta$. In particular, we define the autocorrelation time $t_{\text{aut}}$
as the minimum Monte Carlo time $t$ such that $C(t)/C(0) < 1/e$. We also introduce
another indicator related to the ideal condition~\eqref{eqn:ideal}. Since we expect~\eqref{eqn:frustrated_ineq}
to hold for any algorithm and for each pair of interacting spins on the lattice, we
have
\begin{equation}\label{eqn:sum_ineq}
\frac{1}{4N^2} \sum_{i,j} \left| \thermal{S_i S_j} \right| \le \frac{1}{4N^2} \sum_{i,j}
\thermal{\gamma_{ij}}_{\text{sb}},
\end{equation}
where $i$ and $j$ denote sites on the double-chain lattice. By introducing the mean
cluster size
\[
\Lambda = \frac{1}{2N}\sum_{s=1}^{2N} N_s s^2,
\]
where $N_s$ stands for the number of cluster of size $s$, it is easy to show that
the right-hand side of~\eqref{eqn:sum_ineq} coincides with $\lambda \equiv\frac{1}{2N}\thermal{
\Lambda}_{\text{sb}}$. Therefore,
\begin{equation}\label{eqn:global_ineq}
m^2_{\text{abs}} \le \lambda,
\end{equation}
where $m^2_{\text{abs}}$ denotes the left-hand side of~\eqref{eqn:sum_ineq}. Finally,
as an efficiency measure, we compare $m^2_{\text{abs}}$ and $\lambda$: the more $\lambda$
exceeds $m^2_{\text{abs}}$, the less efficient the algorithm is expected to be, especially
at criticality. Specifically, we focus on $\beta = 1000$ and $N = 1001$, and compare
the results for each algorithm previously introduced. In Fig.~\ref{fig:aut_time} we
plot the autocorrelation time $t_{\text{aut}}$ as a function of $\alpha$ for several
values of $J$. It is worth noting the existence of a critical value $\alpha_c$ where
the autocorrelation time of each algorithm shows a maximum. In particular, this critical
value is common for each clustering algorithm and depends only on $J$. It is worth
specifying that $\alpha_c$ converges to the critical point of the quantum phase transition
in the thermodynamic limit, i.e. $\beta \to \infty$. We observe that the efficiency
depends on the strength of the antiferromagnetic interaction. Specifically, as $J$
increases, the performance of the algorithm LOC4 becomes even worse than the SW procedure.
The reason underlying this behavior could be the fact that only a small part of the
lattice is treated via frustrated spin plaquettes. This argument supports the idea
that even though the LOC6 technique always shows the best performance, it could suffer
the same issue as the LOC4 one. On the other hand, LR3 and LR4 always yield an improvement
in efficiency with respect to SW, although it could be worse than LOC4 and LOC6 if
$J$ is not large. Moreover, the LR3 and LR4 approaches, in which the residual interaction
can be ferromagnetic, generally perform better than the alternative one in which only
part of the lattice is tiled with plaquettes and the residual interaction remains
antiferromagnetic. This behavior is expected and its consequences become evident when
$J = 1$. In that case, for sufficiently large $\alpha$, specifically, slightly above
$\alpha_c$, the optimal interaction always exceeds $\tilde{J}$, and the algorithm
effectively reduces to the SW scheme. This never occurs for larger values of $J$,
for values of $\alpha$ around the transition, where the residual interaction always
remains ferromagnetic in both cases. In Fig.~\ref{fig:mcs} we plot both $\lambda$
and $m^2_{\text{abs}}$ as functions of $\alpha$ for the values of $J$ considered before.
Especially for large antiferromagnetic interactions, the relative distance $|\lambda
-m^2_{\text{abs}}|/m^2_{\text{abs}}$ can become substantial. Nevertheless, this only
affects the autocorrelation time when $\lambda$ becomes consistently non-vanishing,
i.e. around $\alpha_c$.

\section*{Conclusions}
The results presented in the previous section show that, in general and in almost
all cases, the performance of the proposed algorithms is better than that of the SW
algorithm, with a substantial reduction in the autocorrelation time. As already emphasized,
the long-range approach to Eq.~\eqref{eqn:weights} turns out to be more robust, providing
significant improvements in both partitions we considered, namely LR3 and LR4. On
the other hand, the LOC4 method fails for rectangular plaquettes at $J=5$ — and
presumably for $J>5$ — while the LOC6 approach remains the most efficient. Similarly,
we expect the latter to fail only for values of $J$ much larger than the other energy
scales, thus ensuring a substantial improvement across a wide region of the physically
relevant parameter space. The proposed approach can be further generalized: i) through
the numerical method, i.e. by tiling the entire lattice with plaquettes of the same
type, thereby further reducing lattice connectivity; ii) by considering plaquettes
involving a larger number of spins. Nevertheless, both long-range and local approaches
soon become intractable due to the rapid growth in the number of equations and, more
importantly, of unknowns, which makes selecting the optimal solution increasingly
challenging. Finally, one could attempt to extend the proposed framework to the continuous
imaginary-time limit, i.e., $\tau \to 0$. In this case, clusters of spins (super-spins)
would first be constructed solely through the $\Delta$ term, as is commonly done in
continuous-time algorithms. Subsequently, one should consider frustrated plaquettes
of super-spins arising from the remaining long-range ferromagnetic interactions generated
by the bosonic field integration, together with the equal-time antiferromagnetic couplings
among the super-spins. 
The present study opens the way to the efficient Monte Carlo simulation of groups
of quantum spins coupled to a thermal bath. When the interaction between the spins
is antiferromagnetic, the resulting frustration makes the usual approach based on
the Swendsen-Wang algorithm highly inefficient. On the other hand, the approach presented
here strongly mitigates the effects of frustration, reducing considerably autocorrelation
times.

\bibliography{paper}

\end{document}